\documentclass[trackchanges,twocolumn,resetfootnote]{aastex701}
\usepackage{amsmath}
\usepackage{aas_macros}
\usepackage{graphicx}
\usepackage{wrapfig}
\usepackage{comment}
\usepackage{footnote}
\usepackage{physics}

\begin{document}

\title{High-Order Photon Rings around Kerr Naked Singularities}

\author[orcid=0009-0003-3707-5938]{Hina Suzuki}
\email{hinas@arizona.edu}
\affiliation{Steward Observatory and Department of Astronomy, University of Arizona, 933 N. Cherry Ave., Tucson, AZ~85721, USA}
\affiliation{Department of Electrical and Computer Engineering, University of Arizona, 1230 E. Speedway Blvd., Tucson, AZ~85721, USA}

\author[orcid=0000-0002-8131-6730]{Yosuke Mizuno}
\email{mizuno@sjtu.edu.cn}
\affiliation{Tsung-Dao Lee Institute, Shanghai Jiao-Tong University, Shanghai, 1 Lisuo Road, 201210, People's Republic of China}
\affiliation{School of Physics \& Astronomy, Shanghai Jiao-Tong University, Shanghai, 800 Dongchuan Road, 200240, People's Republic of China}
\affiliation{Key Laboratory for Particle Physics, Astrophysics and Cosmology (MOE), Shanghai Key Laboratory for Particle Physics and Cosmology, Shanghai Jiao-Tong University, 800 Dongchuan Road, Shanghai, 200240, China}

\author[orcid=0000-0001-8213-646X]{Akhil Uniyal}
\email{akhil\_uniyal@sjtu.edu.cn}
\affiliation{Tsung-Dao Lee Institute, Shanghai Jiao-Tong University, Shanghai, 1 Lisuo Road, 201210, People's Republic of China}

\author[orcid=0000-0002-4064-0446]{Indu Kalpa Dihingia}
\email{ikd4638@gmail.com}
\affiliation{Tsung-Dao Lee Institute, Shanghai Jiao-Tong University, Shanghai, 1 Lisuo Road, 201210, People's Republic of China}

\author[orcid=0000-0002-7524-5219]{Tintin Nguyen}
\email{tintin.nguyen@cfa.harvard.edu}
\affiliation{Center for Astrophysics $\vert$ Harvard \& Smithsonian, 60 Garden Street, Cambridge, MA 02138, USA}
\author[orcid=0000-0001-6337-6126]{Chi-kwan Chan}
\email{chanc@arizona.edu}
\affiliation{Steward Observatory and Department of Astronomy, University of Arizona, 933 N. Cherry Ave., Tucson, AZ~85721, USA}
\affiliation{Data Science Institute, University of Arizona, 1230 N. Cherry Ave., Tucson, AZ~85721, USA}
\affiliation{Program in Applied Mathematics, University of Arizona, 617 N. Santa Rita, Tucson, AZ~85721, USA}

\begin{abstract}
We present a detailed study of higher-order photon rings of an accreting Kerr naked singularity (KNS) with dimensionless spin parameter $a=1.01$;
i.e., a horizonless, overly spinning compact object.
Motivated by horizon-scale very-long-baseline interferometry (VLBI) including Event Horizon Telescope (EHT) and future missions such as the Black Hole Explorer (BHEX), we analyze image morphology and interferometric visibilities to identify observational signatures that differentiate KNS from Kerr black holes.
We find that higher-order photon rings are tightly concentrated within the nominal ``shadow'' region and that the shadow develops a pronounced gap at sufficiently large observer inclination.
These morphological differences produce measurable deviations in the complex visibilities relative to Kerr black hole predictions.
Our results indicate that photon-ring structure and visibility-domain diagnostics at horizon-resolving baselines can provide a direct observational test of the presence (or absence) of an event horizon and thus offer a concrete avenue to test general relativity with future horizon-scale observations. 
\end{abstract}


\section{Introduction}\label{sec:intro}
The Event Horizon Telescope (EHT) has opened a new era in experimental gravity, delivering the first horizon-scale images of supermassive black holes---M87$^*$ and Sagittarius A$^*$ (Sgr~A$^*$) \citep{2019ApJ...875L...1E,2022ApJ...930L..12E}. These observations offer an unprecedented opportunity to test the prediction of general relativity (GR) in the strong field regime, particularly through the morphology of black hole shadows and associated photon rings \citep{2010ApJ...718..446J, Mizuno:2018lxz, 2020PhRvL.125n1104P, 2023CQGra..40p5007V, Uniyal:2025etal}.

Mainstream researches interpret these images by assuming that M87$^*$ and Sgr~A$^*$ are Kerr black holes (KBH) with event horizons \cite{2019ApJ...875L...6E, 2023arXiv231109484T}.
However, these interpretations are not unique.
Other compact objects, such as Kerr naked singularities (KNSs) can produce similar shadow images, raising the question of whether horizonless spacetime should also be consistent with exciting observations \citep[e.g.,][]{eht2019theory, psaltis2020test, EventHorizonTelescope:2021dqv}. Recently, ~\cite{Olivares:2018abq} showed that a boson star can be distinguished from a black hole by comparing the image of a non-rotating boson star with Schwarzschild and rapidly spinning Kerr black hole. Therefore, it is important to study other compact objects, such as KNS, to understand how their images differ from those of black holes.

Naked singularities arise when the Kerr spin parameter exceeds the Kerr bound ($|a|\leq 1.0$), where $a = J/M^2$ is the spin parameter of a compact object with mass $M$ and its angular momentum $J$. Although traditionally, naked singularities were considered to be nonphysical objects due to the violation of the cosmic censorship conjecture, they have been studied as a candidate object for testing the limits of classical GR and exploring the phenomenology of extreme gravity.

The weak cosmic censorship conjecture proposes that spacetime singularities resulting from gravitational collapse must be hidden from an observer at infinity by event horizons \citep{1969NCimR...1..252P}. If this is true, this would ensure that classical GR remains predictive in physically relevant spacetime. Violation of this principle would imply a fundamental breakdown of predictability in GR. However, multiple naked singularities can solve Einstein's field equations \citep{1968PhRvL..20..878J, 10.1093/ptep/ptw147,PhysRevLett.66.994,2011CQGra..28w5018J, PhysRevLett.118.181101}. Theoretical studies have demonstrated that naked singularities can also arise in dynamical collapse, alternative gravity models, or semiclassical contexts. Their observational signatures---including altered shadow geometry, bright central regions, and enhanced photon ring structure---may offer a route to test this foundational principle empirically.

Prior works have investigated the shadow geometry and lensing signatures of naked singularities, using analytical or idealized ray-tracing models \citep[e.g.,][]{PhysRevD.79.043002, TintinNKS2023ApJ...954...78N, 2024CQGra..41f5004T, 2021CQGra..38h5010K, PhysRevD.80.024042}. These studies have shown that the spins and observational inclination angles impact observable quantities of naked singularities, such as the arc length, angle, curvature radius, and distortion parameter of the shadow. These studies have established that KNSs can produce image structures similar to black holes, but with subtle differences in the morphology, brightness of photon rings, and brightness depression of the inner region. Recently, some general relativistic magnetohydrodynamic (GRMHD) simulations were performed to distinguish the KNS from the KBH~\citep{NKSGRMHD2025ApJ...978...44D}, which further also investigated, along with other naked singularity solutions to know the strong outflow mechanism responsible around such compact objects~\citep{Uniyal:2025hik}. An extensive study has been done to distinguish Reissner-Nordstr\"om (RN) naked singularities from their black hole counterparts~\citep{Mishra:2024bpl, Kluzniak:2024cxm, Cemeljic:2025bqz}. The intensity of the central brightness depressions in EHT images provides another way to distinguish black holes from other types of naked singularities, such as the Janis-Newman-Winicour and Gauss-Bonnet naked singularities \citep{Deliyski_2025}.

High-order photon rings, formed by photons that orbit around the compact objects multiple times, carry detailed information about the spacetime geometry near the photon sphere \citep{PhysRevD.102.124004}. These rings also provide a more sensitive test of spacetime structure than the shadow boundary alone, which can be highly sensitive to the accretion physics \citep{Johnson:2019ljv}. A major opportunity to probe the feature lies with the Black Hole Explorer (BHEX) mission, a space-VLBI designed to resolve high-order photon rings \citep{2024SPIE13092E..2DJ}. At a longer baseline length of $\sim 27G\lambda$ ($\sim 3$ times higher angular resolution than the EHT), BHEX probes the interferometric regime where the first-order sub-ring becomes dominant for a KBH \citep{Johnson:2019ljv}.

To prepare for such high-precision tests of gravity, it is essential to understand the higher-order photon ring structure in both black holes and naked singularities' spacetime. In particular, the interpretation of BHEX observation will require robust forward modeling of horizonless geometries, including realistic emission and plasma dynamics. So far, this regime has remained largely unexplored, especially in the absence of realistic emission models.

In this paper, we present the first detailed study of higher-order photon rings around a KNS with general relativistic radiative transfer (GRRT) using GRMHD simulated data of KNSs. Building on recent progress in modeling accretion flow onto KNSs \citep{NKSGRMHD2025ApJ...978...44D}, we perform ray tracing through a 3D turbulent plasma to construct a synthetic image and isolate the cascaded photon rings. We shall calculate image of a high-order photon ring at various inclination angles. In order to provide observables with VLBI instruments, we plot visibility amplitude against u-v baseline distance. Our results provide predictive templates of high-resolution imaging and contribute to the development of strong-field tests of general relativity using future VLBI instruments.

The outline of the paper is as follows. In Sec.~\ref{sec:2} we introduce the theoretical expectations for photon rings around KNS. In Sec.~\ref{sec:3} we describe the GRRT simulation setup and the initial conditions used for the GRMHD models. In Sec.~\ref{sec:4} we present the higher-order photon rings of KNS and their dependence on observer inclination. In Sec.~\ref{sec:5} we compute and analyze the visibility amplitudes corresponding to these high-order rings. Finally, in Sec.~\ref{sec:6} we summarize our findings and discuss implications for observational tests of horizonless spacetimes.

\section{Theoretical Expectation for Photon ``Rings'' around KNS} \label{sec:2}

\subsection{Critical Curve and Photon Ring}

Below, we review some technical details regarding photon rings to clarify key terminologies and subsequent calculations. The interested readers could refer to the review by \citealt{Lupsasca_2024_photon_ring_intro} for derivations and further details. Kerr spacetime admits several conserved quantities along null geodesics: energy $E$, azimuthal angular momentum $L$, and Carter's constant $C$ \citep{Carter_1968}. Spherical photon orbits (at constant radius $r = \tilde{r}$) are characterized by conserved quantities $\Phi \equiv L_z/E$ and $Q \equiv C/E^2$ in the following set of equations \citep{Teo_2003}:
\begin{align}
    \Phi &= -\frac{\tilde{r}^3 - 3\tilde{r}^2 + a^2 \tilde{r} + a^2}{a(\tilde{r} -1)} \label{eq:Phi} \\ 
    Q &= \frac{\tilde{r}^3(\tilde{r}^3 - 6 \tilde{r}^2 + 9\tilde{r} -4a^2)}{a^2(\tilde{r}-1)^2} \label{eq:Q}
\end{align}
A photon with conserved quantities $\Phi$ and $Q$ can be projected to an image plane at infinity at an inclination angle $i$ from the spin axis with Cartesian coordinates $(x, y)$ (or equivalently $\alpha, \beta$ in \citealt{1973blho.conf..215B}):
\begin{align}
    x &= -\Phi \csc i \label{eq:image_x} \\
    y &= \pm \sqrt{Q + a^2 \cos^2 i} - \Phi^2 \cot^2 i\label{eq:image_y}
\end{align}
Unstable spherical photon orbits around KNSs exist in the range $r_{\rm ms} < r < r_{\rm ph}$, where $r_{\rm ms}$ and $r_{\rm ph}$ are the marginally stable orbital radius and the equatorial retrograde circular orbital radius, respectively \citep{Charbulak:2018wzb}:
\begin{align}
    r_{\rm ms} &= 1 + (a^2 -1)^{1/3} \label{eq:r_ms} \\
    r_{\rm ph} &= 2 + 2 \cosh \left[ \frac{1}{3} \cosh^{-1} (2a^2-1) \right]
    \label{eq:r_ph}
\end{align}
The projection of unstable spherical photon orbits to an observer at infinity through Equations~\ref{eq:Phi}-\ref{eq:r_ph} trace out the \textit{critical curve} on the image plane. For KBH, the critical curve is a closed boundary separating the interior region of photon capture (the black hole \textit{shadow}) and the exterior region of photon escape. For KNS, the critical curve may develop a gap or not exist for higher viewing inclination and spin \citep{Nguyen_2023}. By definition, the critical curve is an infinitely thin region of bound orbits that cannot be detected by our telescope. Equating the critical curve and the black hole shadow to the central brightness depressions in EHT images requires strong astrophysical assumptions on the accretion physics, thus posing limitations in probing spacetime properties and strong-field tests of gravity.

In practice, the observable \textit{photon ring} is an infinite series of self-similar sub-rings formed by nearly bound orbits, where each sub-ring is indexed by its image order $n$, denoting the number of half orbits the photon make around the KBH or KNS \citep{gralla2019shadows}. Each successive sub-ring (increasing image order in $n$) converges exponentially towards the critical curve at the limit of $n \rightarrow \infty$ and becomes exponentially thinner in width \citep{Johnson:2019ljv}, making them prime targets of space-based VLBI. For example, BHEX aims to resolve the first-order ($n = 1$) photon rings of M87* and Sgr A* \citep{2024SPIE13092E..2DJ}.

\subsection{Comparison Between KNS and KBH Photon Rings}

It is expected that the photon rings around KNS would show significant differences from their black hole counterparts because of the non-existence of the event horizon and the reflective nature of the KNS. For KBHs, the observed photon ring remains essentially circular for any spin parameter and observer inclination~\citep{Psaltis:2018xkc}. The spacetime admits both prograde and retrograde unstable photon trajectories, and their projection on the image plane produces a continuous, closed shadow silhouette. Therefore, in the case of black holes, with an event horizon, cascaded photon rings converge into a critical curve, the projection of the unstable photon orbit onto the image plane of the distant stationary observer \citep{PhysRevD.100.024018}.

In contrast, for the case of the KNS, a subset of prograde photon paths confined to the equatorial plane (and nearby spherical orbits) do not continue to form bound, scattering trajectories but instead spiral into the central singularity or escape to infinity by passing through the inside of a singular ring, leaving behind a dark spot or line depending on the inclination angle. On the other hand, the retrograde unstable photon trajectories produce an arc-like image since they can come back to the observer~\citep{Hioki:2009na, Patel:2022vlu}. Now, in the realistic case, because KNS lack an event horizon, the region that would correspond to the black-hole “shadow’’ is expected to be filled by a highly lensed image of the accretion flow rather than remaining dark. Hence, photon rings exterior to the critical curve are expected to resemble those of black holes, but the nominal shadow interior will generally contain lensed emission. \citet{TintinNKS2023ApJ...954...78N, NKSGRMHD2025ApJ...978...44D} have shown that, for sufficiently high spin and near edge-on inclinations, the photon rings can develop an earlier discussed gap, and for larger spin or inclination, the rings may disappear entirely. How emission is lensed through or across this gap remains an open theoretical problem. However, higher-order photon images should behave qualitatively differently because there is no horizon to absorb light; hence, whenever an $n$-order image appears outside the critical curve, a corresponding counterpart is expected to appear inside the curve. We have discussed this in detail in the following sections.


\section{Numerical Setup} \label{sec:3}

This study analyzes the geometry of high-order photon rings using a dataset from 3D GRMHD simulations performed with the \texttt{BHAC} code \citep{Porth2017, 2019A&A...629A..61O} in modified Kerr–Schild coordinates \citep[Details of GRMHD simulation can be found at:][]{NKSGRMHD2025ApJ...978...44D}. For the GRMHD simulation, an ``inflow'' boundary condition has been adopted at the inner boundary of the simulation outside the naked singularity at $r=0.3680\,r_g$. Note that the results from the GRMHD simulations become independent of whether the inner boundary is ``inflow'' or ``reflective'' if the boundary is close enough to the singularity (see Appendix B of \cite{NKSGRMHD2025ApJ...978...44D}). We extract 400 snapshots over the simulation ranging from $t=1000\,t_g - 5000\,t_g$ and compute ray-traced images using a relativistic radiative transfer code, \texttt{RAPTOR} \citep{2018A&A...613A...2B, 2020A&A...641A.126B}.  The source is considered Sgr~A$^*$, having mass $M = 4.14 \times 10^6 \, M_{\odot}$ at distance 8.127 $\mathrm{kpc}$. By choosing $R_{high} = R_{low} = 1$ for simplicity, we assume a single-temperature plasma emission exclusively from thermal synchrotron radiation. We also rescale the mass accretion rate to produce an optically thin image, ensuring that the emission becomes transparent at 230 GHz so that the underlying photon-ring structure can be robustly identified and analyzed in image domain.
We calculated images with the grid resolution in the image plane being $1000 \times 1000$ pixels, and the field of view 40 $r_g^2$ (0.25 $\mathrm{mas}^2$). Figures~\ref{fig:gap} and~\ref{fig:subrings} show time averaged images in 8 $r_g^2$ (0.05 $\mathrm{mas}^2$) for visualization.

For a better comparison, we considered a range of inclination angles of observers placed at $i = [15^\circ, 30^\circ, 45^\circ, 60^\circ, 75^\circ]$. The images are calculated at 230~GHz to match the EHT observing frequency. Each image is normalized and visualized in a quadratic scale for better comparison. Furthermore, we used the publicly available \texttt{eventhorizontelescope/2017-sgra-paper5} pipeline to compute visibility amplitude curves consistent with EHT data products. Sub-ring decomposition is performed by isolating photon trajectories based on their half-winding number around the KNS. We analyze the contribution of each sub-ring to the total image and study its visibility signatures in the $uv$ domain.

\section{High-Order Photon Ring around KNS} \label{sec:4}

\begin{figure}[h]
  \centering
  \includegraphics[width=\columnwidth]{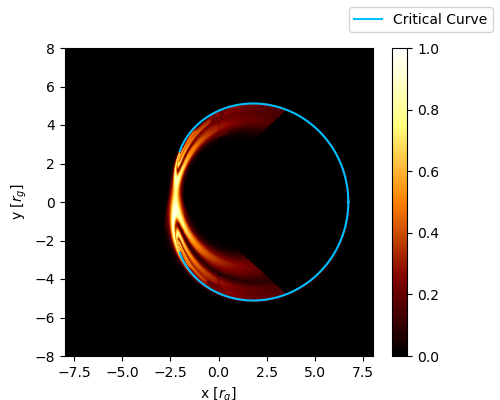}
  \caption{High-Order Photon ring structure around a KNS with spin $a = 1.01$ and a polar inclination angle $i = 60^\circ$, ray-traced from 3D GRMHD simulation data from~\citep{NKSGRMHD2025ApJ...978...44D}. The critical curve is calculated with the analytical calculations (Equations \ref{eq:Phi}-\ref{eq:r_ph}). Multiple cascaded photon rings appear on both sides of the critical curve (blue) due to the absence of an event horizon. Inner rings are thicker and more sparse, connected with outer rings around the equatorial plane. Intensity is normalized and visualized on a quadratic scale to enhance the visibility of the sub-ring. Dark regions on the right side are likely artifacts from the numerical boundary in the simulation and do not affect the analysis of the VLBI observational signature.}
  \label{fig:gap}
\end{figure}

In this section, we computed images of KNS at different observer inclinations and examine how the higher-order photon rings change with inclination. Figure~\ref{fig:gap} shows the higher-order photon rings for KNS at inclination angle $i=60^\circ$. The blue curve denotes the critical curve, analytically computed with Equations \ref{eq:Phi}-\ref{eq:r_ph}. Unlike the nearly circular photon rings seen around KBH, the KNS produces a pronounced, multilayered thick–ring structure. Because there is no event horizon, rings appear both inside and outside the critical curve and connect across the equatorial plane, with the inner cascades appearing broader and more sparsely spaced. For clarity, all intensities are normalized and displayed using a quadratic intensity scaling to enhance the faint, high-order features. To compensate for the numerical limitation along the polar axis in GRRT calculations, we interpolate along a distorted radial coordinate defined with respect to the critical curve. \footnote{
Interpolation was performed along the azimuthal direction $\phi$ on a coordinate system defined by the critical curve. The distorted coordinate were constructed from cartesian $(x,y)$ to distorted polar coordinate $(r, \phi)$ as  $r = \sqrt{x^2+y^2} / r_\mathrm{crit}(\phi)$ and $\phi = \tan^{-1}(y/x)$, where $r_\mathrm{crit}$ is the radius of critical curve as a function of $\phi$. The values were interpolated using the  \texttt{scipy.interpolate.griddata} function with the linear method, which performs piecewise linear interpolation between neighboring data points. 
} 

The dark regions visible on the right of the figure are most likely artifacts caused by the finite inner boundary of the GRMHD simulation, which was imposed for computational reasons. Although the inner boundary can in principle be moved closer to the singularity ($r=0$), the overall dynamics is insensitive to this choice for our case because the KNS is modeled with a reflective inner boundary~\citep{NKSGRMHD2025ApJ...978...44D, Uniyal:2025hik}, therefore, does not impact our VLBI observational signature of the KNS.


\begin{figure*}
  \centering
  \includegraphics[width=\textwidth]{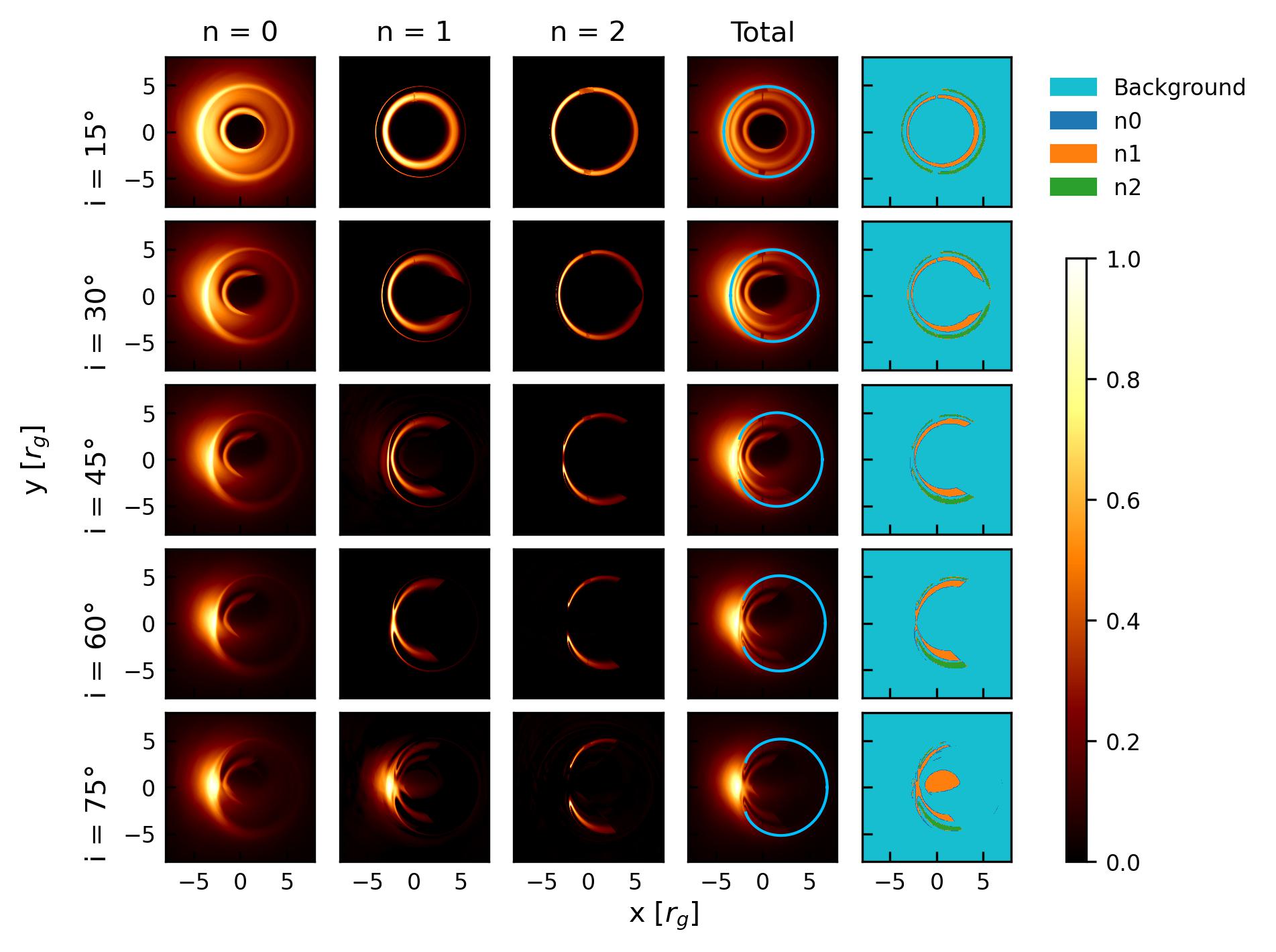}
  \caption{Decomposition of the sub-rings at different inclinations. Two cascaded rings per order of photon ring appear on both sides of the critical curve. For all inclination angles, rings inside the critical curve have a thicker profile than those outside. At the higher inclination angles, high-order photon rings begin to open a gap. The right panels display the spatial contribution of each sub-ring to the total image. Similar to KBH, the intensity decreases as the order of the photon ring increases, but the presence of inner rings introduces additional structure within the shadow region.}
  \label{fig:subrings}
\end{figure*}

Figure~\ref{fig:subrings} shows the decomposition of sub-rings and their contribution to the total image at different inclinations. The right-most panels represent the clear structure of the higher-order rings for each inclination angle. For all inclination angles, we observe two rings on both inside and outside of the critical curve. We find that a gap in the left of a ring structure only appears for higher-order photon rings ($n \geq 2$) and at sufficiently large inclinations ($i>30^{\circ}$), where the gap begins to open up. KNSs are super-spinning ($a>1$), so prograde photons that co-rotate with the spacetime can acquire sufficiently large effective angular momentum that the centrifugal barrier prevents them from maintaining bound (spherical) orbits~\citep{Charbulak:2018wzb}. For a fixed spin, increasing the observer inclination from face-on toward edge-on increases the photon's motion in the $\phi$ direction and thus increases the magnitude of their effective angular momentum. This drives prograde trajectories to become progressively more prograde, and once those trajectories can no longer sustain spherical orbits, they produce a gap in the critical curve at sufficiently large inclinations. In particular, for $a = 1.01$~\citep{TintinNKS2023ApJ...954...78N} indicates that the critical curve begins to open a gap at viewing angles roughly $i \gtrsim 31^{\circ}$. 

The critical curve itself is an unobservable, infinitesimally thin locus, but photons emitted just outside it trace nearly bound trajectories that orbit arbitrarily close to the photon sphere. This behavior is evident in Figure~\ref{fig:subrings}, where the gap that first appears on the critical curve is already manifest in the $n=2$ images. By contrast, $n=1$ photons complete only roughly half an orbit around the KNS and therefore do not experience the same centrifugal barrier-driven removal. Consequently, the $n=1$ sub-ring does not develop a gap. Additionally, because the $n=1$ ring is slightly offset from the critical curve, we do not see a gap in the $n=1$ image for $a=1.01$ at $i=60^\circ$. In any case, whether a gap presents or not, the inner and outer ring components join smoothly at the edge of the critical curve.

\section{VLBI Signature} \label{sec:5}

In the previous section, we discussed how the higher-order rings show a gap with the different inclination angles, which is not present in the case of the BHs. We next examine the visibility-domain signatures of the higher-order rings to quantify their observational appearance. 
VLBI arrays detect a complex visibility $V(\vb{u})$, which is the Fourier transform of the corresponding the Sky image $I(\vb{x})$\citep{2017isra.book.....T}. 

\begin{align}
    V(\vb{u}) = \int I(\vb{x})e^{-2\pi i \vb{u}\cdot \vb{x}}d\vb{x}
\end{align}
, where $\vb{u}$ in ($\lambda$) is baseline vector projected orthogonal onto the line of sight, and $\vb{x}$ in (radians) is image coordinate. For the case of a thin, uniform, circular ring, image and visibility function on the observer plane, in polar coordinates ($r$, $\varphi$), is 
\begin{align}
    I(r, \varphi) & = \frac{1}{\pi d}\delta(r-\frac{d}{2}) \\
    V(u, \varphi_u) & = J_0(\pi du)
\end{align}
Here, $J_m$ are $m$th Bessel functions of first kind. 
\begin{figure*}
    \centering
  \includegraphics[width=\textwidth]{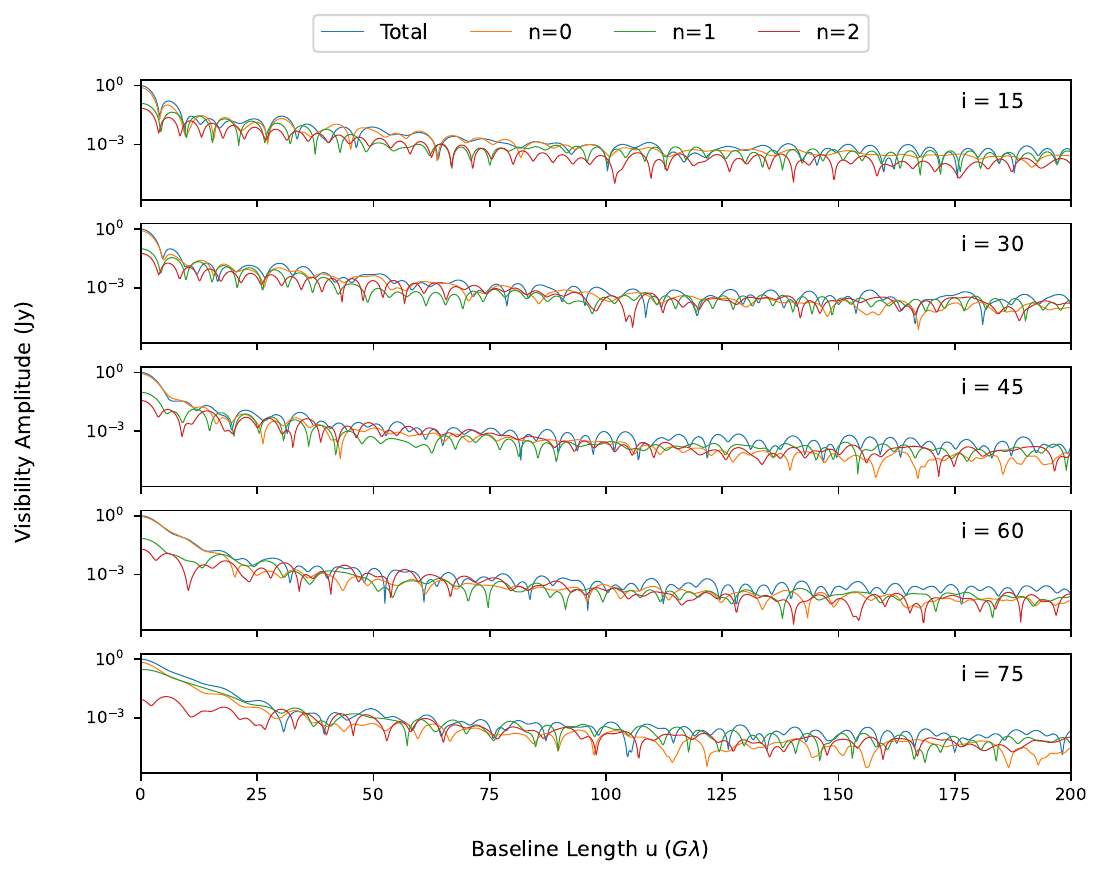}
    \caption{Normalized visibility amplitude of KNS ($a=1.01$) as a function of $(u, v)$-distance for five inclination angles, shown from top to bottom. Each curve represents the amplitude from cascaded sub-rings. The absence of an event horizon produces a thicker emission structure, resulting in similar decay rates of the visibility amplitude across sub-rings compared to the sharper fall-off seen in that of black holes.}
  \label{fig:va}
\end{figure*}
Figure~\ref{fig:va} shows the normalized visibility amplitude (Jy) as a function of baseline length ($G\lambda$) for the decomposed sub-rings and for the full image at different inclinations. 

Visibilities are computed from $0$ to $200$  $G\lambda$. The visibility oscillation frequency reflects the characteristic ring diameter, while the oscillation envelope exhibits only weak decay across photon-ring orders. This slow fall-off arises because the emission in the KNS images is extended rather than concentrated in an infinitely thin ring. Thus, significant power remains at high spatial frequencies even for higher-order sub-rings compared to the BHs~\citep{Wong:2024gph, Capistrano:2025kht}.

\section{Discussion} \label{sec:6}

In this work, we have carried out a systematic study of simulated images of accreting KNSs and contrasted their appearance with that of KBHs. Our analysis identifies several distinct morphological and radiative signatures that arise from the absence of an event horizon, providing potential observational diagnostics that can distinguish horizonless spacetimes.

We identified a cascaded photon-ring structure in which successive photon-ring orders form distinct, layered features. As the inclination angle increases, these rings develop a gap consistent with earlier theoretical predictions; however, the gap emerges only in the higher-order components of the image. This is highly relevant to future space-based VLBI mission such as BHEX, which can only probe the $n=1$ photon ring of M87* and Sgr A*. Even for KNS models where the critical curve (or higher-order sub-rings at $n \geq 2$) opens a gap (e.g, $a = 1.01$, $i \gtrsim 30^{\circ}$ models), they cannot be ruled out by EHT and BHEX observations of a closed photon ring image dominated by lower-order ($n = 0$ and $n = 1$) components. Conclusively ruling out KNS models such as $a = 1.01$, $i \gtrsim 30^{\circ}$ based on whether an observed photon ring image develops a gap or not requires much longer baseline at $\gtrsim 200$ G$\lambda$ (e.g. on the Moon) to probe $n =2$ photon rings.

These morphological differences produce distinctive visibility-domain signatures, suggesting that interferometric observations at horizon-scale resolution could potentially distinguish KNS from KBH images. In particular, the cascaded photon-ring structure and the inclination-dependent gap are likely to induce measurable modulations in the visibility amplitude accessible to next-generation VLBI facilities (e.g., BHEX). The absence of the event horizon produces a thicker emission structure due to the inner ring features. The width of each KNS sub-ring becomes thinner for each successive order at a slower rate than KBH sub-rings, resulting in similar decay rates of the visibility amplitude across sub-rings, in contrast to the sharper fall-off observed in that of KBHs. As a consequence, whereas each progressively longer baseline regime is dominated by the emission of a single sub-ring at a particular image order for KBHs \citep{Johnson:2019ljv}, different KNSs sub-rings are not cleanly separated in the visibility response at longer baselines, as seen in Figure \ref{fig:va}.

The extension of this work includes surveying a wider range of spin parameters, incorporating polarized radiative transfer to characterize polarization signatures near the gap and inside the critical curve, and studying time-dependent behavior and correlations. A complementary approach with isolating emission from the nominal shadow region and analyzing residual should help quantify the fundamental morphological differences between KNS and KBH models. Finally, mapping the image morphologies across spin and inclination will allow a direct comparison between our GRMHD-based ray-tracing results and analytic predictions, providing a more comprehensive test of horizonless spacetime phenomenology. We planned to do it in the future.

Therefore, by linking GRMHD-based numerical simulations with analytic geodesic structures, this work provides a unified theoretical framework for distinguishing black holes from naked singularities through their observable image features.

\begin{acknowledgments}
Y.M., I.K.D., and A.U. are supported by the National Key R\&D Program of China (Grant No.\,2023YFE0101200), the National Natural Science Foundation of China (Grant No.\,12273022), the Research Fund for Excellent International PhD Students (grant No. W2442004) and the Shanghai Municipality orientation program of Basic Research for International Scientists (Grant No.\,22JC1410600). I.K.D. acknowledges the TDLI postdoctoral fellowship for financial support.
\end{acknowledgments}

\bibliographystyle{aasjournal}
\bibliography{refs}

@ARTICLE{NKSGRMHD2025ApJ...978...44D,
       author = {{Dihingia}, Indu K. and {Uniyal}, Akhil and {Mizuno}, Yosuke},
        title = "{Distinguishability of a Naked Singularity from a Black Hole in Dynamics and Radiative Signatures}",
      journal = {\apj},
         year = 2025,
        month = jan,
       volume = {978},
       number = {1},
          eid = {44},
        pages = {44},
          doi = {10.3847/1538-4357/ad9600},
archivePrefix = {arXiv},
       eprint = {2410.13406},
 primaryClass = {astro-ph.HE},
       adsurl = {https://ui.adsabs.harvard.edu/abs/2025ApJ...978...44D}
}

@ARTICLE{TintinNKS2023ApJ...954...78N,
       author = {{Nguyen}, Bao and {Christian}, Pierre and {Chan}, Chi-kwan},
        title = "{Shadow Geometry of Kerr Naked Singularities}",
      journal = {\apj},
         year = 2023,
        month = sep,
       volume = {954},
       number = {1},
          eid = {78},
        pages = {78},
          doi = {10.3847/1538-4357/ace697},
archivePrefix = {arXiv},
       eprint = {2302.08094},
 primaryClass = {astro-ph.HE},
       adsurl = {https://ui.adsabs.harvard.edu/abs/2023ApJ...954...78N}
}

@ARTICLE{2022ApJ...930L..12E,
       author = {{Event Horizon Telescope Collaboration} and {Akiyama}, Kazunori and {Alberdi}, Antxon and {Alef}, Walter and {Algaba}, Juan Carlos and {Anantua}, Richard and {Asada}, Keiichi and {Azulay}, Rebecca and {Bach}, Uwe and {Baczko}, Anne-Kathrin and {Ball}, David and {Balokovi{\'c}}, Mislav and {Barrett}, John and {Baub{\"o}ck}, Michi and {Benson}, Bradford A. and {Bintley}, Dan and {Blackburn}, Lindy and {Blundell}, Raymond and {Bouman}, Katherine L. and {Bower}, Geoffrey C. and {Boyce}, Hope and {Bremer}, Michael and {Brinkerink}, Christiaan D. and {Brissenden}, Roger and {Britzen}, Silke and {Broderick}, Avery E. and {Broguiere}, Dominique and {Bronzwaer}, Thomas and {Bustamante}, Sandra and {Byun}, Do-Young and {Carlstrom}, John E. and {Ceccobello}, Chiara and {Chael}, Andrew and {Chan}, Chi-kwan and {Chatterjee}, Koushik and {Chatterjee}, Shami and {Chen}, Ming-Tang and {Chen}, Yongjun and {Cheng}, Xiaopeng and {Cho}, Ilje and {Christian}, Pierre and {Conroy}, Nicholas S. and {Conway}, John E. and {Cordes}, James M. and {Crawford}, Thomas M. and {Crew}, Geoffrey B. and {Cruz-Osorio}, Alejandro and {Cui}, Yuzhu and {Davelaar}, Jordy and {De Laurentis}, Mariafelicia and {Deane}, Roger and {Dempsey}, Jessica and {Desvignes}, Gregory and {Dexter}, Jason and {Dhruv}, Vedant and {Doeleman}, Sheperd S. and {Dougal}, Sean and {Dzib}, Sergio A. and {Eatough}, Ralph P. and {Emami}, Razieh and {Falcke}, Heino and {Farah}, Joseph and {Fish}, Vincent L. and {Fomalont}, Ed and {Ford}, H. Alyson and {Fraga-Encinas}, Raquel and {Freeman}, William T. and {Friberg}, Per and {Fromm}, Christian M. and {Fuentes}, Antonio and {Galison}, Peter and {Gammie}, Charles F. and {Garc{\'\i}a}, Roberto and {Gentaz}, Olivier and {Georgiev}, Boris and {Goddi}, Ciriaco and {Gold}, Roman and {G{\'o}mez-Ruiz}, Arturo I. and {G{\'o}mez}, Jos{\'e} L. and {Gu}, Minfeng and {Gurwell}, Mark and {Hada}, Kazuhiro and {Haggard}, Daryl and {Haworth}, Kari and {Hecht}, Michael H. and {Hesper}, Ronald and {Heumann}, Dirk and {Ho}, Luis C. and {Ho}, Paul and {Honma}, Mareki and {Huang}, Chih-Wei L. and {Huang}, Lei and {Hughes}, David H. and {Ikeda}, Shiro and {Impellizzeri}, C.~M. Violette and {Inoue}, Makoto and {Issaoun}, Sara and {James}, David J. and {Jannuzi}, Buell T. and {Janssen}, Michael and {Jeter}, Britton and {Jiang}, Wu and {Jim{\'e}nez-Rosales}, Alejandra and {Johnson}, Michael D. and {Jorstad}, Svetlana and {Joshi}, Abhishek V. and {Jung}, Taehyun and {Karami}, Mansour and {Karuppusamy}, Ramesh and {Kawashima}, Tomohisa and {Keating}, Garrett K. and {Kettenis}, Mark and {Kim}, Dong-Jin and {Kim}, Jae-Young and {Kim}, Jongsoo and {Kim}, Junhan and {Kino}, Motoki and {Koay}, Jun Yi and {Kocherlakota}, Prashant and {Kofuji}, Yutaro and {Koch}, Patrick M. and {Koyama}, Shoko and {Kramer}, Carsten and {Kramer}, Michael and {Krichbaum}, Thomas P. and {Kuo}, Cheng-Yu and {La Bella}, Noemi and {Lauer}, Tod R. and {Lee}, Daeyoung and {Lee}, Sang-Sung and {Leung}, Po Kin and {Levis}, Aviad and {Li}, Zhiyuan and {Lico}, Rocco and {Lindahl}, Greg and {Lindqvist}, Michael and {Lisakov}, Mikhail and {Liu}, Jun and {Liu}, Kuo and {Liuzzo}, Elisabetta and {Lo}, Wen-Ping and {Lobanov}, Andrei P. and {Loinard}, Laurent and {Lonsdale}, Colin J. and {Lu}, Ru-Sen and {Mao}, Jirong and {Marchili}, Nicola and {Markoff}, Sera and {Marrone}, Daniel P. and {Marscher}, Alan P. and {Mart{\'\i}-Vidal}, Iv{\'a}n and {Matsushita}, Satoki and {Matthews}, Lynn D. and {Medeiros}, Lia and {Menten}, Karl M. and {Michalik}, Daniel and {Mizuno}, Izumi and {Mizuno}, Yosuke and {Moran}, James M. and {Moriyama}, Kotaro and {Moscibrodzka}, Monika and {M{\"u}ller}, Cornelia and {Mus}, Alejandro and {Musoke}, Gibwa and {Myserlis}, Ioannis and {Nadolski}, Andrew and {Nagai}, Hiroshi and {Nagar}, Neil M. and {Nakamura}, Masanori and {Narayan}, Ramesh and {Narayanan}, Gopal and {Natarajan}, Iniyan and {Nathanail}, Antonios and {Fuentes}, Santiago Navarro and {Neilsen}, Joey and {Neri}, Roberto and {Ni}, Chunchong and {Noutsos}, Aristeidis and {Nowak}, Michael A. and {Oh}, Junghwan and {Okino}, Hiroki and {Olivares}, H{\'e}ctor and {Ortiz-Le{\'o}n}, Gisela N. and {Oyama}, Tomoaki and {{\"O}zel}, Feryal and {Palumbo}, Daniel C.~M. and {Paraschos}, Georgios Filippos and {Park}, Jongho and {Parsons}, Harriet and {Patel}, Nimesh and {Pen}, Ue-Li and {Pesce}, Dominic W. and {Pi{\'e}tu}, Vincent and {Plambeck}, Richard and {PopStefanija}, Aleksandar and {Porth}, Oliver and {P{\"o}tzl}, Felix M. and {Prather}, Ben and {Preciado-L{\'o}pez}, Jorge A. and {Psaltis}, Dimitrios},
        title = "{First Sagittarius A* Event Horizon Telescope Results. I. The Shadow of the Supermassive Black Hole in the Center of the Milky Way}",
      journal = {\apjl},
         year = 2022,
        month = may,
       volume = {930},
       number = {2},
          eid = {L12},
        pages = {L12},
          doi = {10.3847/2041-8213/ac6674},
       adsurl = {https://ui.adsabs.harvard.edu/abs/2022ApJ...930L..12E}
}

@article{gralla2019shadows,
  title={Black hole shadows, photon rings, and lensing rings},
  author={Gralla, Samuel E and Holz, Daniel E and Wald, Robert M},
  journal={Physical Review D},
  volume={100},
  number={2},
  pages={024018},
  year={2019},
  doi={10.1103/PhysRevD.100.024018},
  eprint={1906.00873},
  archivePrefix={arXiv}
}

@article{eht2019theory,
  title={First M87 Event Horizon Telescope Results. V. Physical Origin of the Asymmetric Ring},
  author={Event Horizon Telescope Collaboration},
  journal={The Astrophysical Journal Letters},
  volume={875},
  number={1},
  pages={L5},
  year={2019},
  doi={10.3847/2041-8213/ab0f43},
  eprint={1906.11242},
  archivePrefix={arXiv}
}

@article{psaltis2020test,
  title={A General Relativistic Null Hypothesis Test with Event Horizon Telescope Observations of the Black Hole Shadow in M87},
  author={Psaltis, Dimitrios and others},
  journal={Physical Review Letters},
  volume={125},
  number={14},
  pages={141104},
  year={2020},
  doi={10.1103/PhysRevLett.125.141104},
  eprint={2008.11701},
  archivePrefix={arXiv}
}

@ARTICLE{2019ApJ...875L...1E,
       author = {{Event Horizon Telescope Collaboration} and {Akiyama}, Kazunori and {Alberdi}, Antxon and {Alef}, Walter and {Asada}, Keiichi and {Azulay}, Rebecca and {Baczko}, Anne-Kathrin and {Ball}, David and {Balokovi{\'c}}, Mislav and {Barrett}, John and {Bintley}, Dan and {Blackburn}, Lindy and {Boland}, Wilfred and {Bouman}, Katherine L. and {Bower}, Geoffrey C. and {Bremer}, Michael and {Brinkerink}, Christiaan D. and {Brissenden}, Roger and {Britzen}, Silke and {Broderick}, Avery E. and {Broguiere}, Dominique and {Bronzwaer}, Thomas and {Byun}, Do-Young and {Carlstrom}, John E. and {Chael}, Andrew and {Chan}, Chi-kwan and {Chatterjee}, Shami and {Chatterjee}, Koushik and {Chen}, Ming-Tang and {Chen}, Yongjun and {Cho}, Ilje and {Christian}, Pierre and {Conway}, John E. and {Cordes}, James M. and {Crew}, Geoffrey B. and {Cui}, Yuzhu and {Davelaar}, Jordy and {De Laurentis}, Mariafelicia and {Deane}, Roger and {Dempsey}, Jessica and {Desvignes}, Gregory and {Dexter}, Jason and {Doeleman}, Sheperd S. and {Eatough}, Ralph P. and {Falcke}, Heino and {Fish}, Vincent L. and {Fomalont}, Ed and {Fraga-Encinas}, Raquel and {Freeman}, William T. and {Friberg}, Per and {Fromm}, Christian M. and {G{\'o}mez}, Jos{\'e} L. and {Galison}, Peter and {Gammie}, Charles F. and {Garc{\'\i}a}, Roberto and {Gentaz}, Olivier and {Georgiev}, Boris and {Goddi}, Ciriaco and {Gold}, Roman and {Gu}, Minfeng and {Gurwell}, Mark and {Hada}, Kazuhiro and {Hecht}, Michael H. and {Hesper}, Ronald and {Ho}, Luis C. and {Ho}, Paul and {Honma}, Mareki and {Huang}, Chih-Wei L. and {Huang}, Lei and {Hughes}, David H. and {Ikeda}, Shiro and {Inoue}, Makoto and {Issaoun}, Sara and {James}, David J. and {Jannuzi}, Buell T. and {Janssen}, Michael and {Jeter}, Britton and {Jiang}, Wu and {Johnson}, Michael D. and {Jorstad}, Svetlana and {Jung}, Taehyun and {Karami}, Mansour and {Karuppusamy}, Ramesh and {Kawashima}, Tomohisa and {Keating}, Garrett K. and {Kettenis}, Mark and {Kim}, Jae-Young and {Kim}, Junhan and {Kim}, Jongsoo and {Kino}, Motoki and {Koay}, Jun Yi and {Koch}, Patrick M. and {Koyama}, Shoko and {Kramer}, Michael and {Kramer}, Carsten and {Krichbaum}, Thomas P. and {Kuo}, Cheng-Yu and {Lauer}, Tod R. and {Lee}, Sang-Sung and {Li}, Yan-Rong and {Li}, Zhiyuan and {Lindqvist}, Michael and {Liu}, Kuo and {Liuzzo}, Elisabetta and {Lo}, Wen-Ping and {Lobanov}, Andrei P. and {Loinard}, Laurent and {Lonsdale}, Colin and {Lu}, Ru-Sen and {MacDonald}, Nicholas R. and {Mao}, Jirong and {Markoff}, Sera and {Marrone}, Daniel P. and {Marscher}, Alan P. and {Mart{\'\i}-Vidal}, Iv{\'a}n and {Matsushita}, Satoki and {Matthews}, Lynn D. and {Medeiros}, Lia and {Menten}, Karl M. and {Mizuno}, Yosuke and {Mizuno}, Izumi and {Moran}, James M. and {Moriyama}, Kotaro and {Moscibrodzka}, Monika and {M{\"u}ller}, Cornelia and {Nagai}, Hiroshi and {Nagar}, Neil M. and {Nakamura}, Masanori and {Narayan}, Ramesh and {Narayanan}, Gopal and {Natarajan}, Iniyan and {Neri}, Roberto and {Ni}, Chunchong and {Noutsos}, Aristeidis and {Okino}, Hiroki and {Olivares}, H{\'e}ctor and {Ortiz-Le{\'o}n}, Gisela N. and {Oyama}, Tomoaki and {{\"O}zel}, Feryal and {Palumbo}, Daniel C.~M. and {Patel}, Nimesh and {Pen}, Ue-Li and {Pesce}, Dominic W. and {Pi{\'e}tu}, Vincent and {Plambeck}, Richard and {PopStefanija}, Aleksandar and {Porth}, Oliver and {Prather}, Ben and {Preciado-L{\'o}pez}, Jorge A. and {Psaltis}, Dimitrios and {Pu}, Hung-Yi and {Ramakrishnan}, Venkatessh and {Rao}, Ramprasad and {Rawlings}, Mark G. and {Raymond}, Alexander W. and {Rezzolla}, Luciano and {Ripperda}, Bart and {Roelofs}, Freek and {Rogers}, Alan and {Ros}, Eduardo and {Rose}, Mel and {Roshanineshat}, Arash and {Rottmann}, Helge and {Roy}, Alan L. and {Ruszczyk}, Chet and {Ryan}, Benjamin R. and {Rygl}, Kazi L.~J. and {S{\'a}nchez}, Salvador and {S{\'a}nchez-Arguelles}, David and {Sasada}, Mahito and {Savolainen}, Tuomas and {Schloerb}, F. Peter and {Schuster}, Karl-Friedrich and {Shao}, Lijing and {Shen}, Zhiqiang and {Small}, Des and {Sohn}, Bong Won and {SooHoo}, Jason and {Tazaki}, Fumie and {Tiede}, Paul and {Tilanus}, Remo P.~J. and {Titus}, Michael and {Toma}, Kenji and {Torne}, Pablo and {Trent}, Tyler and {Trippe}, Sascha and {Tsuda}, Shuichiro and {van Bemmel}, Ilse and {van Langevelde}, Huib Jan and {van Rossum}, Daniel R. and {Wagner}, Jan and {Wardle}, John and {Weintroub}, Jonathan and {Wex}, Norbert and {Wharton}, Robert and {Wielgus}, Maciek and {Wong}, George N. and {Wu}, Qingwen and {Young}, Ken and {Young}, Andr{\'e}},
        title = "{First M87 Event Horizon Telescope Results. I. The Shadow of the Supermassive Black Hole}",
      journal = {\apjl},
         year = 2019,
        month = apr,
       volume = {875},
       number = {1},
          eid = {L1},
        pages = {L1},
          doi = {10.3847/2041-8213/ab0ec7},
archivePrefix = {arXiv},
       eprint = {1906.11238},
 primaryClass = {astro-ph.GA},
       adsurl = {https://ui.adsabs.harvard.edu/abs/2019ApJ...875L...1E}
}

@article{PhysRevD.79.043002,
  title = {Apparent shape of super-spinning black holes},
  author = {Bambi, Cosimo and Freese, Katherine},
  journal = {Phys. Rev. D},
  volume = {79},
  issue = {4},
  pages = {043002},
  numpages = {7},
  year = {2009},
  month = {Feb},
  publisher = {American Physical Society},
  doi = {10.1103/PhysRevD.79.043002},
  url = {https://link.aps.org/doi/10.1103/PhysRevD.79.043002}
}

@article{Nguyen_2023,
doi = {10.3847/1538-4357/ace697},
url = {https://dx.doi.org/10.3847/1538-4357/ace697},
year = {2023},
month = {aug},
publisher = {The American Astronomical Society},
volume = {954},
number = {1},
pages = {78},
author = {Nguyen, Bao and Christian, Pierre and Chan, Chi-kwan},
title = {Shadow Geometry of Kerr Naked Singularities},
journal = {The Astrophysical Journal}
}

@INPROCEEDINGS{2024SPIE13092E..2DJ,
       author = {{Johnson}, Michael D. and {Akiyama}, Kazunori and {Baturin}, Rebecca and {Bilyeu}, Bryan and {Blackburn}, Lindy and {Boroson}, Don and {C{\'a}rdenas-Avenda{\~n}o}, Alejandro and {Chael}, Andrew and {Chan}, Chi-kwan and {Chang}, Dominic and {Cheimets}, Peter and {Chou}, Cathy and {Doeleman}, Sheperd S. and {Farah}, Joseph and {Galison}, Peter and {Gamble}, Ronald and {Gammie}, Charles F. and {Gelles}, Zachary and {G{\'o}mez}, Jos{\'e} L. and {Gralla}, Samuel E. and {Grimes}, Paul and {Gurvits}, Leonid I. and {Hadar}, Shahar and {Haworth}, Kari and {Hada}, Kazuhiro and {Hecht}, Michael H. and {Honma}, Mareki and {Houston}, Janice and {Hudson}, Ben and {Issaoun}, Sara and {Jia}, He and {Jorstad}, Svetlana and {Kauffman}, Jens and {Kovalev}, Yuri Y. and {Kurczynski}, Peter and {Lafon}, Robert E. and {Lupsasca}, Alexandru and {Lehmensiek}, Robert and {Ma}, Chung-Pei and {Marrone}, Daniel P. and {Marscher}, Alan P. and {Melnick}, Gary and {Narayan}, Ramesh and {Niinuma}, Kotaro and {Noble}, Scott C. and {Palmer}, Eric J. and {Palumbo}, Daniel C.~M. and {Paritsky}, Lenny and {Peretz}, Eliad and {Pesce}, Dominic and {Plavin}, Alexander and {Quataert}, Eliot and {Rana}, Hannah and {Ricarte}, Angelo and {Roelofs}, Freek and {Shtyrkova}, Katia and {Sinclair}, Laura C. and {Small}, Jeffrey and {Kumara}, Sridharan Tirupati and {Srinivasan}, Ranjani and {Strominger}, Andrew and {Tiede}, Paul and {Tong}, Edward and {Wang}, Jade and {Weintroub}, Jonathan and {Wielgus}, Maciek and {Wong}, George},
        title = "{The Black Hole Explorer: motivation and vision}",
    booktitle = {Space Telescopes and Instrumentation 2024: Optical, Infrared, and Millimeter Wave},
         year = 2024,
       editor = {{Coyle}, Laura E. and {Matsuura}, Shuji and {Perrin}, Marshall D.},
       series = {Society of Photo-Optical Instrumentation Engineers (SPIE) Conference Series},
       volume = {13092},
        month = aug,
          eid = {130922D},
        pages = {130922D},
          doi = {10.1117/12.3019835},
archivePrefix = {arXiv},
       eprint = {2406.12917},
 primaryClass = {astro-ph.IM},
       adsurl = {https://ui.adsabs.harvard.edu/abs/2024SPIE13092E..2DJ}
}

@article{PhysRevD.100.024018,
  title = {Black hole shadows, photon rings, and lensing rings},
  author = {Gralla, Samuel E. and Holz, Daniel E. and Wald, Robert M.},
  journal = {Phys. Rev. D},
  volume = {100},
  issue = {2},
  pages = {024018},
  numpages = {14},
  year = {2019},
  month = {Jul},
  publisher = {American Physical Society},
  doi = {10.1103/PhysRevD.100.024018},
  url = {https://link.aps.org/doi/10.1103/PhysRevD.100.024018}
}

@article{PhysRevD.102.124004,
  title = {The shape of the black hole photon ring: A precise test of strong-field general relativity},
  author = {Gralla, Samuel E. and Lupsasca, Alexandru and Marrone, Daniel P.},
  journal = {Phys. Rev. D},
  volume = {102},
  issue = {12},
  pages = {124004},
  numpages = {20},
  year = {2020},
  month = {Dec},
  publisher = {American Physical Society},
  doi = {10.1103/PhysRevD.102.124004},
  url = {https://link.aps.org/doi/10.1103/PhysRevD.102.124004}
}

@ARTICLE{1969NCimR...1..252P,
       author = {{Penrose}, Roger},
        title = "{Gravitational Collapse: the Role of General Relativity}",
      journal = {Nuovo Cimento Rivista Serie},
         year = 1969,
        month = jan,
       volume = {1},
        pages = {252},
       adsurl = {https://ui.adsabs.harvard.edu/abs/1969NCimR...1..252P}
}

@ARTICLE{1968PhRvL..20..878J,
       author = {{Janis}, Allen I. and {Newman}, Ezra T. and {Winicour}, Jeffrey},
        title = "{Reality of the Schwarzschild Singularity}",
      journal = {\prl},
         year = 1968,
        month = apr,
       volume = {20},
       number = {16},
        pages = {878-880},
          doi = {10.1103/PhysRevLett.20.878},
       adsurl = {https://ui.adsabs.harvard.edu/abs/1968PhRvL..20..878J}
}

@article{10.1093/ptep/ptw147,
    author = {Mizuno, Ryosuke and Ohashi, Seiju and Shiromizu, Tetsuya},
    title = {Violation of cosmic censorship in the gravitational collapse of a dust cloud in five dimensions},
    journal = {Progress of Theoretical and Experimental Physics},
    volume = {2016},
    number = {10},
    pages = {103E03},
    year = {2016},
    month = {10},
    issn = {2050-3911},
    doi = {10.1093/ptep/ptw147},
    url = {https://doi.org/10.1093/ptep/ptw147},
    eprint = {https://academic.oup.com/ptep/article-pdf/2016/10/103E03/17640315/ptw147.pdf},
}

@article{PhysRevLett.66.994,
  title = {Formation of naked singularities: The violation of cosmic censorship},
  author = {Shapiro, Stuart L. and Teukolsky, Saul A.},
  journal = {Phys. Rev. Lett.},
  volume = {66},
  issue = {8},
  pages = {994--997},
  numpages = {0},
  year = {1991},
  month = {Feb},
  publisher = {American Physical Society},
  doi = {10.1103/PhysRevLett.66.994},
  url = {https://link.aps.org/doi/10.1103/PhysRevLett.66.994}
}

@ARTICLE{2011CQGra..28w5018J,
       author = {{Joshi}, Pankaj S. and {Malafarina}, Daniele and {Narayan}, Ramesh},
        title = "{Equilibrium configurations from gravitational collapse}",
      journal = {Classical and Quantum Gravity},
         year = 2011,
        month = dec,
       volume = {28},
       number = {23},
          eid = {235018},
        pages = {235018},
          doi = {10.1088/0264-9381/28/23/235018},
archivePrefix = {arXiv},
       eprint = {1106.5438},
 primaryClass = {gr-qc},
       adsurl = {https://ui.adsabs.harvard.edu/abs/2011CQGra..28w5018J}
}

@article{PhysRevLett.118.181101,
  title = {Violating the Weak Cosmic Censorship Conjecture in Four-Dimensional Anti--de Sitter Space},
  author = {Crisford, Toby and Santos, Jorge E.},
  journal = {Phys. Rev. Lett.},
  volume = {118},
  issue = {18},
  pages = {181101},
  numpages = {5},
  year = {2017},
  month = {May},
  publisher = {American Physical Society},
  doi = {10.1103/PhysRevLett.118.181101},
  url = {https://link.aps.org/doi/10.1103/PhysRevLett.118.181101}
}

@ARTICLE{2019A&A...629A..61O,
       author = {{Olivares}, Hector and {Porth}, Oliver and {Davelaar}, Jordy and {Most}, Elias R. and {Fromm}, Christian M. and {Mizuno}, Yosuke and {Younsi}, Ziri and {Rezzolla}, Luciano},
        title = "{Constrained transport and adaptive mesh refinement in the Black Hole Accretion Code}",
      journal = {\aap},
         year = 2019,
        month = sep,
       volume = {629},
          eid = {A61},
        pages = {A61},
          doi = {10.1051/0004-6361/201935559},
archivePrefix = {arXiv},
       eprint = {1906.10795},
 primaryClass = {astro-ph.HE},
       adsurl = {https://ui.adsabs.harvard.edu/abs/2019A&A...629A..61O}
}

@Article{Porth2017,
author={Porth, Oliver
and Olivares, Hector
and Mizuno, Yosuke
and Younsi, Ziri
and Rezzolla, Luciano
and Moscibrodzka, Monika
and Falcke, Heino
and Kramer, Michael},
title={The black hole accretion code},
journal={Computational Astrophysics and Cosmology},
year={2017},
month={May},
day={03},
volume={4},
number={1},
pages={1},
issn={2197-7909},
doi={10.1186/s40668-017-0020-2},
url={https://doi.org/10.1186/s40668-017-0020-2}
}

@ARTICLE{2018A&A...613A...2B,
       author = {{Bronzwaer}, T. and {Davelaar}, J. and {Younsi}, Z. and {Mo{\'s}cibrodzka}, M. and {Falcke}, H. and {Kramer}, M. and {Rezzolla}, L.},
        title = "{RAPTOR. I. Time-dependent radiative transfer in arbitrary spacetimes}",
      journal = {\aap},
         year = 2018,
        month = may,
       volume = {613},
          eid = {A2},
        pages = {A2},
          doi = {10.1051/0004-6361/201732149},
archivePrefix = {arXiv},
       eprint = {1801.10452},
 primaryClass = {astro-ph.HE},
       adsurl = {https://ui.adsabs.harvard.edu/abs/2018A&A...613A...2B}
}

@ARTICLE{2020A&A...641A.126B,
       author = {{Bronzwaer}, T. and {Younsi}, Z. and {Davelaar}, J. and {Falcke}, H.},
        title = "{RAPTOR. II. Polarized radiative transfer in curved spacetime}",
      journal = {\aap},
         year = 2020,
        month = sep,
       volume = {641},
          eid = {A126},
        pages = {A126},
          doi = {10.1051/0004-6361/202038573},
archivePrefix = {arXiv},
       eprint = {2007.03045},
 primaryClass = {astro-ph.HE},
       adsurl = {https://ui.adsabs.harvard.edu/abs/2020A&A...641A.126B}
}

@INPROCEEDINGS{1973blho.conf..215B,
       author = {{Bardeen}, J.~M.},
        title = "{Timelike and null geodesics in the Kerr metric.}",
    booktitle = {Black Holes (Les Astres Occlus)},
         year = 1973,
       editor = {{Dewitt}, C. and {Dewitt}, B.~S.},
        month = jan,
        pages = {215-239},
       adsurl = {https://ui.adsabs.harvard.edu/abs/1973blho.conf..215B}
}

@article{Johnson:2019ljv,
    author = "Johnson, Michael D. and others",
    title = "{Universal interferometric signatures of a black hole{\textquoteright}s photon ring}",
    eprint = "1907.04329",
    archivePrefix = "arXiv",
    primaryClass = "astro-ph.IM",
    doi = "10.1126/sciadv.aaz1310",
    journal = "Sci. Adv.",
    volume = "6",
    number = "12",
    pages = "eaaz1310",
    year = "2020"
}

@ARTICLE{2024CQGra..41f5004T,
       author = {{Tavlayan}, Aydin and {Tekin}, Bayram},
        title = "{Instability of a Kerr-type naked singularity due to light and matter accretion and its shadow}",
      journal = {Classical and Quantum Gravity},
         year = 2024,
        month = mar,
       volume = {41},
       number = {6},
          eid = {065004},
        pages = {065004},
          doi = {10.1088/1361-6382/ad2318},
archivePrefix = {arXiv},
       eprint = {2301.13751},
 primaryClass = {gr-qc},
       adsurl = {https://ui.adsabs.harvard.edu/abs/2024CQGra..41f5004T}
}

@ARTICLE{2021CQGra..38h5010K,
       author = {{Kumar}, Rahul and {Ghosh}, Sushant G.},
        title = "{Photon ring structure of rotating regular black holes and no-horizon spacetimes}",
      journal = {Classical and Quantum Gravity},
         year = 2021,
        month = apr,
       volume = {38},
       number = {8},
          eid = {085010},
        pages = {085010},
          doi = {10.1088/1361-6382/abdd48},
archivePrefix = {arXiv},
       eprint = {2004.07501},
 primaryClass = {gr-qc},
       adsurl = {https://ui.adsabs.harvard.edu/abs/2021CQGra..38h5010K}
}

@article{PhysRevD.80.024042,
  title = {Measurement of the Kerr spin parameter by observation of a compact object's shadow},
  author = {Hioki, Kenta and Maeda, Kei-ichi},
  journal = {Phys. Rev. D},
  volume = {80},
  issue = {2},
  pages = {024042},
  numpages = {9},
  year = {2009},
  month = {Jul},
  publisher = {American Physical Society},
  doi = {10.1103/PhysRevD.80.024042},
  url = {https://link.aps.org/doi/10.1103/PhysRevD.80.024042}
}

@ARTICLE{2020PhRvL.125n1104P,
       author = {{Psaltis}, Dimitrios and {Medeiros}, Lia and {Christian}, Pierre and {{\"O}zel}, Feryal and {Akiyama}, Kazunori and {Alberdi}, Antxon and {Alef}, Walter and {Asada}, Keiichi and {Azulay}, Rebecca and {Ball}, David and {Balokovi{\'c}}, Mislav and {Barrett}, John and {Bintley}, Dan and {Blackburn}, Lindy and {Boland}, Wilfred and {Bower}, Geoffrey C. and {Bremer}, Michael and {Brinkerink}, Christiaan D. and {Brissenden}, Roger and {Britzen}, Silke and {Broguiere}, Dominique and {Bronzwaer}, Thomas and {Byun}, Do-Young and {Carlstrom}, John E. and {Chael}, Andrew and {Chan}, Chi-kwan and {Chatterjee}, Shami and {Chatterjee}, Koushik and {Chen}, Ming-Tang and {Chen}, Yongjun and {Cho}, Ilje and {Conway}, John E. and {Cordes}, James M. and {Crew}, Geoffrey B. and {Cui}, Yuzhu and {Davelaar}, Jordy and {De Laurentis}, Mariafelicia and {Deane}, Roger and {Dempsey}, Jessica and {Desvignes}, Gregory and {Dexter}, Jason and {Eatough}, Ralph P. and {Falcke}, Heino and {Fish}, Vincent L. and {Fomalont}, Ed and {Fraga-Encinas}, Raquel and {Friberg}, Per and {Fromm}, Christian M. and {Gammie}, Charles F. and {Garc{\'\i}a}, Roberto and {Gentaz}, Olivier and {Goddi}, Ciriaco and {G{\'o}mez}, Jos{\'e} L. and {Gu}, Minfeng and {Gurwell}, Mark and {Hada}, Kazuhiro and {Hesper}, Ronald and {Ho}, Luis C. and {Ho}, Paul and {Honma}, Mareki and {Huang}, Chih-Wei L. and {Huang}, Lei and {Hughes}, David H. and {Inoue}, Makoto and {Issaoun}, Sara and {James}, David J. and {Jannuzi}, Buell T. and {Janssen}, Michael and {Jiang}, Wu and {Jimenez-Rosales}, Alejandra and {Johnson}, Michael D. and {Jorstad}, Svetlana and {Jung}, Taehyun and {Karami}, Mansour and {Karuppusamy}, Ramesh and {Kawashima}, Tomohisa and {Keating}, Garrett K. and {Kettenis}, Mark and {Kim}, Jae-Young and {Kim}, Junhan and {Kim}, Jongsoo and {Kino}, Motoki and {Koay}, Jun Yi and {Koch}, Patrick M. and {Koyama}, Shoko and {Kramer}, Michael and {Kramer}, Carsten and {Krichbaum}, Thomas P. and {Kuo}, Cheng-Yu and {Lauer}, Tod R. and {Lee}, Sang-Sung and {Li}, Yan-Rong and {Li}, Zhiyuan and {Lindqvist}, Michael and {Lico}, Rocco and {Liu}, Jun and {Liu}, Kuo and {Liuzzo}, Elisabetta and {Lo}, Wen-Ping and {Lobanov}, Andrei P. and {Lonsdale}, Colin and {Lu}, Ru-Sen and {Mao}, Jirong and {Markoff}, Sera and {Marrone}, Daniel P. and {Marscher}, Alan P. and {Mart{\'\i}-Vidal}, Iv{\'a}n and {Matsushita}, Satoki and {Mizuno}, Yosuke and {Mizuno}, Izumi and {Moran}, James M. and {Moriyama}, Kotaro and {Moscibrodzka}, Monika and {M{\"u}ller}, Cornelia and {Musoke}, Gibwa and {Mus Mej{\'\i}as}, Alejandro and {Nagai}, Hiroshi and {Nagar}, Neil M. and {Narayan}, Ramesh and {Narayanan}, Gopal and {Natarajan}, Iniyan and {Neri}, Roberto and {Noutsos}, Aristeidis and {Okino}, Hiroki and {Olivares}, H{\'e}ctor and {Oyama}, Tomoaki and {Palumbo}, Daniel C.~M. and {Park}, Jongho and {Patel}, Nimesh and {Pen}, Ue-Li and {Pi{\'e}tu}, Vincent and {Plambeck}, Richard and {PopStefanija}, Aleksandar and {Prather}, Ben and {Preciado-L{\'o}pez}, Jorge A. and {Ramakrishnan}, Venkatessh and {Rao}, Ramprasad and {Rawlings}, Mark G. and {Raymond}, Alexander W. and {Ripperda}, Bart and {Roelofs}, Freek and {Rogers}, Alan and {Ros}, Eduardo and {Rose}, Mel and {Roshanineshat}, Arash and {Rottmann}, Helge and {Roy}, Alan L. and {Ruszczyk}, Chet and {Ryan}, Benjamin R. and {Rygl}, Kazi L.~J. and {S{\'a}nchez}, Salvador and {S{\'a}nchez-Arguelles}, David and {Sasada}, Mahito and {Savolainen}, Tuomas and {Schloerb}, F. Peter and {Schuster}, Karl-Friedrich and {Shao}, Lijing and {Shen}, Zhiqiang and {Small}, Des and {Sohn}, Bong Won and {SooHoo}, Jason and {Tazaki}, Fumie and {Tilanus}, Remo P.~J. and {Titus}, Michael and {Torne}, Pablo and {Trent}, Tyler and {Traianou}, Efthalia and {Trippe}, Sascha and {van Bemmel}, Ilse and {van Langevelde}, Huib Jan and {van Rossum}, Daniel R. and {Wagner}, Jan and {Wardle}, John and {Ward-Thompson}, Derek and {Weintroub}, Jonathan and {Wex}, Norbert and {Wharton}, Robert and {Wielgus}, Maciek and {Wong}, George N. and {Wu}, Qingwen and {Yoon}, Doosoo and {Young}, Andr{\'e} and {Young}, Ken and {Younsi}, Ziri and {Yuan}, Feng and {Yuan}, Ye-Fei and {Zhao}, Shan-Shan and {EHT Collaboration}},
        title = "{Gravitational Test beyond the First Post-Newtonian Order with the Shadow of the M87 Black Hole}",
      journal = {\prl},
         year = 2020,
        month = oct,
       volume = {125},
       number = {14},
          eid = {141104},
        pages = {141104},
          doi = {10.1103/PhysRevLett.125.141104},
archivePrefix = {arXiv},
       eprint = {2010.01055},
 primaryClass = {gr-qc},
       adsurl = {https://ui.adsabs.harvard.edu/abs/2020PhRvL.125n1104P}
}

@ARTICLE{2023CQGra..40p5007V,
       author = {{Vagnozzi}, Sunny and {Roy}, Rittick and {Tsai}, Yu-Dai and {Visinelli}, Luca and {Afrin}, Misba and {Allahyari}, Alireza and {Bambhaniya}, Parth and {Dey}, Dipanjan and {Ghosh}, Sushant G. and {Joshi}, Pankaj S. and {Jusufi}, Kimet and {Khodadi}, Mohsen and {Walia}, Rahul Kumar and {{\"O}vg{\"u}n}, Ali and {Bambi}, Cosimo},
        title = "{Horizon-scale tests of gravity theories and fundamental physics from the Event Horizon Telescope image of Sagittarius A (*)}",
      journal = {Classical and Quantum Gravity},
         year = 2023,
        month = aug,
       volume = {40},
       number = {16},
          eid = {165007},
        pages = {165007},
          doi = {10.1088/1361-6382/acd97b},
archivePrefix = {arXiv},
       eprint = {2205.07787},
 primaryClass = {gr-qc},
       adsurl = {https://ui.adsabs.harvard.edu/abs/2023CQGra..40p5007V}
}

@ARTICLE{2023arXiv231109484T,
       author = {{The Event Horizon Telescope Collaboration}},
        title = "{First Sagittarius A* Event Horizon Telescope Results. VI: Testing the Black Hole Metric}",
      journal = {arXiv e-prints},
         year = 2023,
        month = nov,
          eid = {arXiv:2311.09484},
        pages = {arXiv:2311.09484},
          doi = {10.48550/arXiv.2311.09484},
archivePrefix = {arXiv},
       eprint = {2311.09484},
 primaryClass = {astro-ph.HE},
       adsurl = {https://ui.adsabs.harvard.edu/abs/2023arXiv231109484T}
}

@article{Hioki:2009na,
    author = "Hioki, Kenta and Maeda, Kei-ichi",
    title = "{Measurement of the Kerr Spin Parameter by Observation of a Compact Object's Shadow}",
    eprint = "0904.3575",
    archivePrefix = "arXiv",
    primaryClass = "astro-ph.HE",
    reportNumber = "WU-AP-299-09",
    doi = "10.1103/PhysRevD.80.024042",
    journal = "Phys. Rev. D",
    volume = "80",
    pages = "024042",
    year = "2009"
}

@article{Charbulak:2018wzb,
    author = "Charbul{\'a}k, Daniel and Stuchl{\'\i}k, Zden{\v{e}}k",
    title = "{Spherical photon orbits in the field of Kerr naked singularities}",
    eprint = "1811.02648",
    archivePrefix = "arXiv",
    primaryClass = "gr-qc",
    doi = "10.1140/epjc/s10052-018-6336-5",
    journal = "Eur. Phys. J. C",
    volume = "78",
    number = "11",
    pages = "879",
    year = "2018"
}

\end{document}